\documentclass[aps,prb,preprint,12pt]{revtex4-1}  

\usepackage{amsmath}  
\usepackage{amsfonts} 
\usepackage{graphicx} 
\usepackage{color}

\newcommand{\be}[1]{\begin{equation}\label{#1}}
\newcommand{\ee}{\end{equation}}
\newcommand{\bea}[1]{\begin{eqnarray}\label{#1}}
\newcommand{\eea}{\end{eqnarray}}
\newcommand{\no}{\nonumber \\}
\newcommand{\Fig}[1]{Fig.(\ref{#1})}

\newcommand{\Eq}[1]{Eq.(\ref{#1})}

\newcommand{\Sec}[1]{Section~\ref{#1}}
\newcommand{\bsub}{\begin{subequations}}
\newcommand{\esub}{\end{subequations}}
\newcommand{\bwt}{\begin{widetext}}
\newcommand{\ewt}{\end{widetext}}

\usepackage{color}

\def\trm#1{\textrm{#1}}
\def\tit#1{\textit{#1}}
\def\tbf#1{\textbf{#1}}

\def\a0{{\alpha_0}}

\def\da0{{\dot{\alpha}_0}}

\def\myoverDefn#1#2{\hbox{\space \raise-2mm\hbox{$\textstyle{#1} \atop \scriptstyle{#2}$} }}
\def\defn{\overset{\textrm{def}}{=}}

\def\a{{\alpha}}

\def\dag{\dagger}

\def\rp{r_{P}}
\def\rp2{r_{p}^{2}}

\def\A{\mathcal{A}}
\newcommand{\half}{\frac{1}{2}}
\newcommand{\thalf}{\tfrac{1}{2}}

\newcommand{\ket}[1]{|#1\rangle}
\newcommand{\bra}[1]{\langle #1|}

\newcommand{\IP}[2]{\langle {#1} | {#2} \rangle}

\begin{document}

\title{The Hong-Ou-Mandel effect is really odd}
\author{Paul M. Alsing}\email{corresponding author: palsing@albany.edu}
\affiliation{University at Albany-SUNY, Albany, NY 12222, USA}
\author{Richard J. Birrittella}
\affiliation{Booz Allen Hamilton, 8283 Greensboro Drive, McLean, VA 22102, USA}
\affiliation{Air Force Research Laboratory, Information Directorate, 525 Brooks Rd, Rome, NY, 13411, USA}
\author{Christopher C. Gerry}
\affiliation{Department of Physics and Astronomy, The City University of New York, Bronx, New York, 10468-1589, USA}
\author{Jihane Mimih}
\affiliation{Department of Electrical and Computer Engineering, Naval Postgraduate School, 1 University Circle, Monterey, California 93943, USA }
\author{Peter L. Knight}
\affiliation{Blackett Laboratory, Imperial College, London SW72AZ, UK}




\date{\today}

\begin{abstract}
When quantum state amplitudes interfere, surprising non-classical features emerge which emphasis the roles of indistinguishability and discreteness in quantum mechanics. A famous example in quantum optics is the Hong Ou Mandel interference effect,\cite{HOM:1987} a major ingredient in current quantum information processing using photonics. Traditionally the HOM features interference between amplitudes for two one-photon number states. Surprisingly, interference can be manifested when one amplitude represents that most classical of light field states, the coherent state, provided the partner state is non-classical (eg a single photon state or an odd photon number state). Imposing such nonclassical features on an otherwise classical state is the focus of this article.
Recently, the HOM effect has been generalized to the multi-photon case, termed the extended HOM effect by the authors.\cite{eHOM_PRA:2022} The implication of the extended HOM effect is that if an odd parity state, comprising only odd numbers of photons, enters one input port of a  50:50 beam splitter, then regardless of the state entering the other input port, be it pure or mixed, there will no output coincident counts.
In this work, we explain the extended HOM as arising from a sequence of pairwise HOM-like complete destructive interferences occurring simultaneously in the multicomponent amplitude for the output coincidence counts. We first demonstrate this diagrammatically in order to build physical intuition, before developing a general  analytical proof.
We then examine the case of a single photon interacting with a coherent state (and idealized laser), and consider prospects for experimental detection by including the effect of imperfect detection efficiency.
This work highlights the importance of the non-classicality of light, and in particular the interference effects stemming from the discreteness of photon quanta.

\end{abstract}

\maketitle 

\section{Introduction}\label{sec:Intro} 
The generation of two-mode entangled states of light can be accomplished by mixing nonclassical single-mode states of light at a beam splitter. \cite{Kim:2002} 
The process that gives rise to such two-mode states of light via beam splitting is known 
as multiphoton interference 
\cite{Ou:1996,Ou:2007,Ou_Book:2017},  
and serves as a critical element in several applications including quantum optical interferometry \cite{Pan:2012}, and quantum state engineering where beam splitters and conditional measurements are utilized to perform post-selection techniques such as photon subtraction \cite{Dakna:1997,Carranza:2012,Magana-Loaiza:2019}, photon addition  \cite{Dakna:1998}, and photon catalysis. \cite{Lvovsky:2002, Bartley:2012, Birrittella:2018}   

In spite of its name, ``multiphoton interference" does not involve the interference of photons. Rather, as has been emphasized by Glauber \cite{Glauber:1995}, 
it is always the addition of the quantum amplitudes (themselves being complex numbers) associated with these states that give rise to interference effects. 
In effect, quantum amplitudes behave as square roots of probabilities, multiplied by complex phases, allowing for the possibility of        interferences.
The amplitudes to be added are those associated with different paths (or processes) to obtain a given final output state. Thus, the term “multiphoton interference” must be understood to mean “interference with states containing numerous photons.” The canonical example of this kind of interference is what has come to be known as the Hong-Ou-Mandel (HOM) effect \cite{HOM:1987}, 
which is a two-photon (destructive) interference effect wherein single photons 
in either of the output beams 
of a lossless 50:50 beam splitter emerge together (probabilistically). 
(For an extensive historical review of HOM effect and its applications, see the recent review article by Bouchard \tit{et al.} \cite{Bouchard:2021}).
Detectors placed at each of the output ports will yield no simultaneous coincident clicks.
That is, the input state $\ket{1,1}_{12}$ (on modes 1 and 2),   results in the output state  $\tfrac{1}{\sqrt{2}}\left( \ket{2,0}_{12} + \ket{0,2}_{12}\right)$. The absence of the  $\ket{1,1}_{12}$ in the output is due to the complete destructive interference between the component quantum amplitudes of the  two processes (both photons transmitted, or both reflected) that potentially would lead to the state  being in the output.
The essence of this effect from an experimental point of view is that the joint probability $P_{12}(1,1)$  for detecting one photon in each of output beams vanishes, i.e.   $P_{12}(1,1)=0$.

\begin{figure*}[th]
\begin{center}
\begin{tabular}{ccccc}
\hspace{-0.65in}
\includegraphics[width=1.85in,height=1.35in]{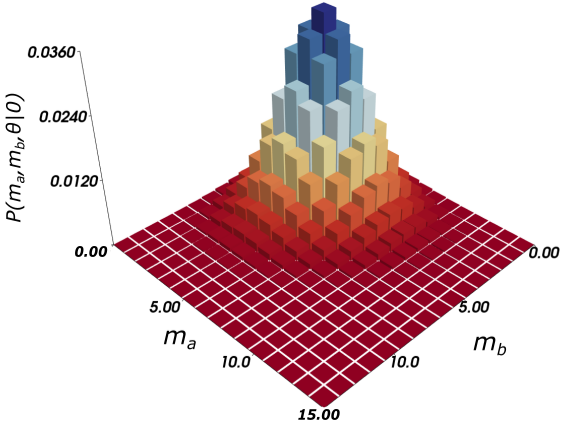}       & {\hspace{2em}}  &
\includegraphics[width=1.85in,height=1.35in]{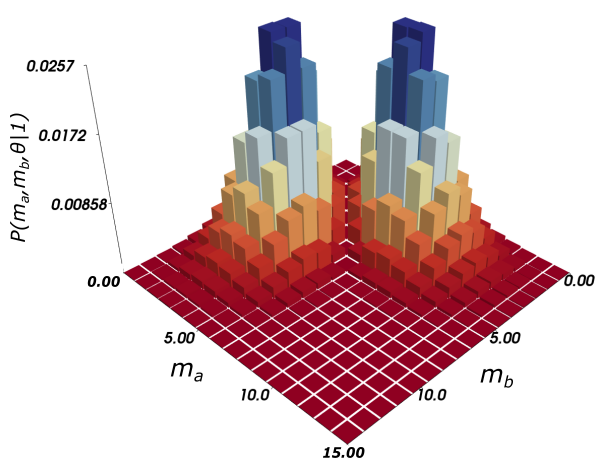}   &{\hspace{2em}}  &
\includegraphics[width=1.85in,height=1.35in]{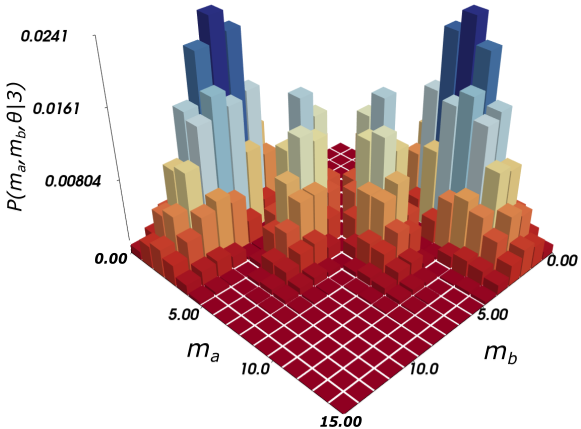} \\
\hspace{-0.65in}
\includegraphics[width=1.85in,height=1.35in]{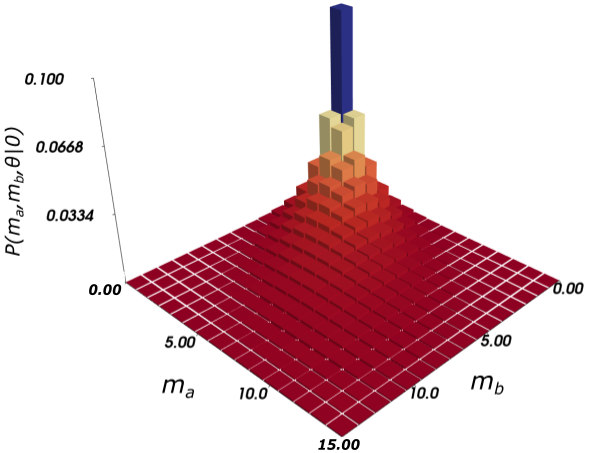}       & {\hspace{2em}}  &
\includegraphics[width=1.85in,height=1.35in]{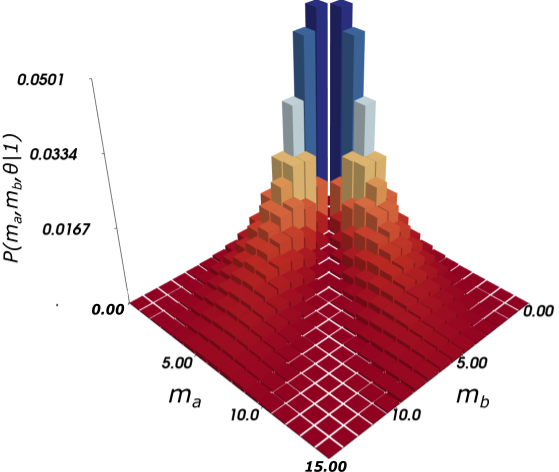}   &{\hspace{2em}}  &
\includegraphics[width=1.85in,height=1.35in]{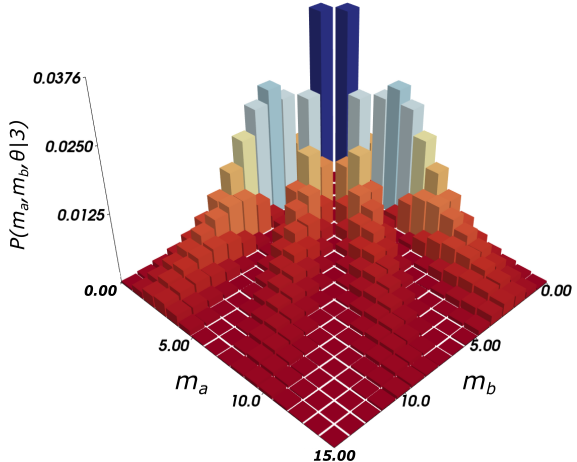}   
\end{tabular}
\end{center} 
\caption{Joint output probability 
$P_{12}(m_a, m_b|n)$ to measure $m_a$ photons in mode-1 and $m_b$ photons in mode-2
from a 50:50 beam splitter
for input  Fock number states $\ket{n}_1$ in mode-1, for $n=\{0,1,3\}$ (top row, left to right),
 and  an input coherent state $\ket{\beta}_2$ in mode-2, with mean number of photons 
with  $\bar{n}_2 =9$.  A central nodal line of zeros for inputs $\ket{n,\beta}_{12}$ 
is observed for odd $n=\{1,3\}$ indicating destructive interference of coincidence 
detection on all output dual Fock states $\ket{m',m'}_{12}$.
No central nodal line is observed for input states with even $n$ (not shown, except for vacuum mode $n=0$), indicating non-zero coincidence detection.
(bottom row) Same as top row, but now with the coherent state mode-2 input state replaced by a mixed thermal state $\rho^{\trm{thermal}}_2$ of average photon number $\bar{n}_2 =9$.
}
\label{fig:eHOM:fig2:fig3:left:column:1photon:CS}    
\end{figure*}

Recently, the authors in \cite{eHOM_PRA:2022} have shown a multi-photon extension of the HOM effect which they termed the extended HOM effect, in which complete destructive interference of the quantum amplitudes for coincident detection output states 
$\ket{\tfrac{n+m}{2}, \tfrac{n+m}{2}}_{12}$
for a balanced (50:50) lossless beam splitter
occurs for any Fock (number) state  input $\ket{n,m}_{12}$  when  $n$ and $m$ are both \tit{odd}, but does not occur when $n$ and $m$ are both \tit{even} (clearly their is no possibility for coincidence detection if $(n,m)$ are either (even,odd) or (odd,even)). Since the  input states $\ket{n,m}_{12}$ form a complete orthonormal dual basis for all bipartite states (both pure and mixed), the extended HOM effect implies that for any odd parity state (comprised of only an odd number of photons) entering the beam splitter input port 1, then regardless of the state entering input port 2, either pure or mixed, there will be a line of zeros, a central nodal line, down the diagonal of the joint output probability for coincidence detection. 
This is illustrated in  \Fig{fig:eHOM:fig2:fig3:left:column:1photon:CS}  showing the joint output probability $P_{12}(m_a, m_b|n)$ to detect $m_a$ photons in output port 1 and $m_b$ photons in output port 2, given that a Fock state $\ket{n}_1$ enters input port 1, for $n=0,1,2,3$ photons. In the top row, the input into port 2 is a coherent state , e.g. an idealized laser, of mean number of photons $\bar{n}_2 = |\beta|^2=9$. The central nodal line in the second and third figures (from left to right) is clearly visible \cite{eHOM:PNCs:note}. 
The bottom row of  \Fig{fig:eHOM:fig2:fig3:left:column:1photon:CS} is similar to the top row, except now the input coherent state is replaced by a mixed thermal state $\rho^{\trm{thermal}}$, again of mean photon number 
$\bar{n}_2 =9$, illustrating the universality of the extended HOM effect.  

The case of the input of a single photon and a coherent state $\ket{1,\beta}_{12}$ was discussed over the years by many authors, most notably Ou (of HOM fame) in 1996 \cite{Ou:1996} (and in subsequent books) \cite{Ou:2007,Ou_Book:2017}, and by Birrittella Mimih and Gerry \cite{BMG:2012} in 2012, but never fully explored as in \cite{eHOM_PRA:2022} (which also examined  non-balanced 
lossless beam splitter configurations as well).  
In his book ``Quantum Optics for Experimentalists," \cite{Ou_Book:2017}, Ou coins the term "Generalized HOM effect," (Chapter 8.3.2) in which judicious choices of the transmission coefficient (for a non-balanced lossless BS) can lead to destructive interference on chosen non-diagonal coincidence states (vs the balanced lossless beam splitter and universality of the central nodal line discussed in this work).
%
The early work of 
Lai, Bu\u{z}ek and Knight (LBK) \cite{Lai:1991}  looked at the beam splitter transformation on dual FS inputs to a fiber-coupler BS, including scattering losses (due to sidewall roughness).
The authors reported: ``\tit{If the same number state enters both ports of the coupler, the probability of finding an odd number of photons at either of the output ports vanishes for a particular choice of the coupler length}." The authors did not explicitly mention that this length is the one appropriate for a 50:50 fiber-coupler beam splitter, which is most certainly the case. 
The work that comes closest to nearly addressing the extended HOM effect was that by Campos, Saleh and Teich \cite{Campos:1989} in their extensive study of the $SU(2)$ properties of a lossless beam splitter. Once again, conditions were discussed for obtaining isolated zeros in the joint output probability distribution, for interesting special cases, but not in all generality.


The outline of this paper is as follows;
In \Sec{sec:BS} we consider the operator beam splitter transformation matrix from input to output modes, and the various forms used in the literature, but whose particular choice/representation does not effect the results of the HOM or extended HOM effects.
In \Sec{sec:Scattering:Amplitudes} we consider the HOM and extended HOM effect for dual Fock state inputs $\ket{n,m}_{12}$ solely in terms of scattering diagrams, keeping track of how input photons are scattered (transmitted or reflected) into output modes, and the $(-1)$ signs they encounter. 
Examination of these diagrams for various dual Fock input states builds up intuition form developing an analytic derivation for arbitrary dual Fock inputs $\ket{n,m}_{12}$, presented at the end of the section.
In \Sec{sec:1:beta:input} we examine the  prospects for an experimental realization of the extended HOM effect on the $\ket{1,1}_{12}$ output coincident state, for an input  consisting of  a single photon Fock state in mode-1 and a coherent state (idealized laser)  in mode-2, by including the effects of finite (imperfect) detection efficiency.
%
 %
 In \Sec{sec:Conclusion} we present a summary and conclusion of our results.
\section{The role of the beam splitter }\label{sec:BS}
The role of a lossless beam splitter has been discussed widely by many authors over the years in many textbooks and articles, \cite{Loudon:2000,Galvez_BS:2002, Skaar:2004, Agarwal:2013,Ou_Book:2017,Gerry_Knight:2023}.
\begin{figure*}[th]
\begin{center}
\includegraphics[width=3.15in,height=2.25in]{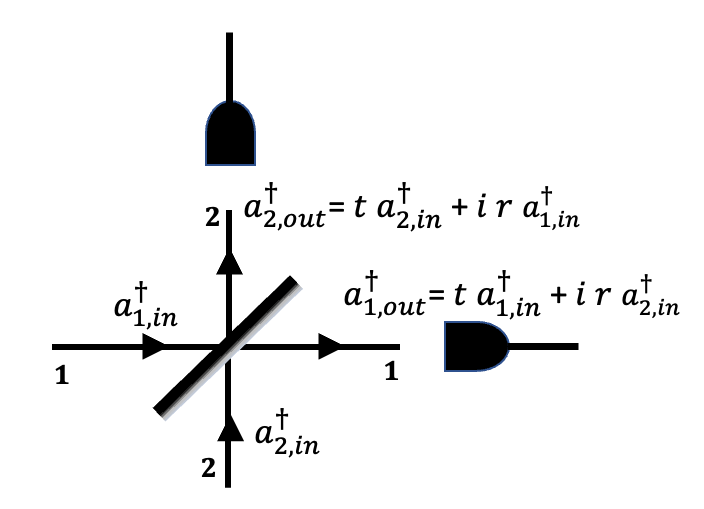}
\end{center} 
\caption{Transformation of  the mode-1 and mode-2 photon creation number operators on a lossless beam splitter (BS), with transmission coefficient $t$ and reflection coefficient $r$ (both real) such that 
$t^2 + r^2\equiv T+R=1.$ 
}
\label{fig:eHOM:BS:schematic_t_ir}    
\end{figure*}
A lossless beam splitter is defined by its representation of the effect of the unitary transformation $U$ of the bosonic  input mode creation  operators $a^\dag_{1,in}, a^\dag_{2,in}$ to the output modes $a^\dag_{1,out}, a^\dag_{2,out}$ via\cite{Skaar:2004}
 $a^\dag_{i,out} = U\,a^\dag_{i,in}\,U^\dag \equiv \sum_{j} S_{ij}\,a^\dag_{j,in}\,$, for  $\;i,j\in\{1,2\} $. Here, the subscripts $1$ and $2$ label the two input/output ports (distinguishable modes) of the BS.
 Unitarity of the transformation between  input modes to output modes is required to ensure that the boson commutation relations of the photon input modes $[a_{i,in}, a^\dag_{j,in}] = \delta_{ij}$ are also satisfied by the output modes $[a_{i,out}, a^\dag_{j,out}] = \delta_{ij}$.

The boson creation operators raise the number of photons in the Fock number state (FS) basis 
$\ket{n}$ by one according to the usual quantum mechanical harmonic oscillator rules (dropping subscripts labels) $a^\dag\ket{n} = \sqrt{n+1}\,\ket{n+1}$. The corresponding annihilation of photons (lowering the occupation number by one) is given by the adjoint operation $a\ket{n} = \sqrt{n}\,\ket{n-1}$. 
In general, the output creation operators can be expressed as a linear combination of the input creation operators, which we express as
\bsub
\bea{aout:intermsof:ain}
a^\dag_{1,out} &=& t\, \,a^\dag_{1,in} + i r\,a^\dag_{2,in}, \label{aout:intermsof:ain:line1}\\
a^\dag_{2,out} &=& i r \,a^\dag_{1,in} + t\,a^\dag_{2,in}, \label{aout:intermsof:ain:line2}
\eea
\esub
where $t$ is the transmission amplitude through the BS, and $r$ is the reflection amplitude (both taken as real), such that $t^2+r^2 \equiv T+R=1$, as shown in \Fig{fig:eHOM:BS:schematic_t_ir}.
For a 50:50 beam splitter, one has $t=r=\tfrac{1}{\sqrt{2}}$. 

Note that in \Eq{aout:intermsof:ain:line1} and \Eq{aout:intermsof:ain:line2} we have chosen to use a (complex) symmetric version of the transformation matrix $S$ defined by 
$a^\dag_{i,out} \defn \sum_{j} S_{ij}\,a^\dag_{j,in}$,
commonly found in the literature and textbooks \cite{Agarwal:2013,Gerry_Knight:2023}.
From \Eq{aout:intermsof:ain:line1} and \Eq{aout:intermsof:ain:line2} we see that the \tit{scattering amplitude} $S_{ij}$ represents the amplitude for a \tit{single} mode-$j$ input photon to scatter (i.e. transmit or reflect) into the output mode-$i$\cite{Skaar:2004}. Thus, our choice above has $S_{11}=S_{22} = t$ and
$S_{12}=S_{21} = i r$, as shown as the leftmost matrix representation in \Eq{S:various:forms}. 
%
\be{S:various:forms}
\hspace{-0.3in}
\textit{representations of\;}
S \rightarrow
%
%
\left[
\begin{array}{cc}
t & i r\\
i r & t
\end{array}
\right], \quad 
\left[
\begin{array}{cc}
t & r\\
 r & -t
\end{array}
\right],\quad 
\left[
\begin{array}{cc}
t & -r\\
r & t
\end{array}
\right], \quad 
%
%
\begin{array}{c}
t= \cos(\theta/2)\\
 r= \sin(\theta/2)
\end{array}, \quad
0\le \theta\le \pi.
\ee

However, the unitarity of $U$, and hence $S$, allows for other possible unitarily equivalent choices (also found in the literature) including the asymmetric middle matrix representation in \Eq{S:various:forms} as used by Skaar\cite{Skaar:2004} appropriate, for example, if mode-1 propagates in a  high index of refraction material, while mode-2 propagates in a low-index material.
Also commonly employed is the (real) anti-symmetric representation\cite{eHOM_PRA:2022} shown as the rightmost matrix in \Eq{S:various:forms}.
In this latter case, the minus sign arises from a $\pi$-phase shift  experienced by light reflected at an interface where the index of refraction changes from low to high\cite{Born_and_Wolf:1986,Jackson:1999,Hecht:2002}.
The commonly employed  choice we have made in  \Eq{aout:intermsof:ain:line1} and \Eq{aout:intermsof:ain:line2} distributes this later $\pi$-phase shift equitably as $\pi/2$ at both beam splitter interfaces, and as an example would be appropriate for a  symmetric beam splitter formed by bringing two strands of fiber optics close together.

The particular choice of $S$ does not effect the outcome of the destructive interference of the quantum amplitudes, as long as $S$ is unitary (though different choices of $S$ correspond to different physical realizations of the particular beam splitter employed, and different relative phases in the output state of the beam splitter).
This point is discussed more fully in the Appendix.

\section{Multi-photon Scattering Amplitudes for Fock State/Fock state inputs}\label{sec:Scattering:Amplitudes}
In this section we consider illustrative and informative multi-photon scattering amplitudes for dual Fock inputs $\ket{n,m}_{12}$. We concentrate in this section on dual Fock inputs, since they form a complete orthonormal basis to construct any bipartite pure state input $\ket{\psi}_{12} = \sum_{n,m}\,c_{nm}\,\ket{n,m}_{12}$, or mixed state input  $\rho_{12} = \sum_{n,m, n',m'}\,\rho_{nm, n'm'}\,\ket{n,m}_{12}\bra{n',m'}$ to the 50:50 BS.
For each input state $\ket{n,m}_{12}$, we consider the total amplitude for the coincident output state 
$\ket{\tfrac{n+m}{2}, \tfrac{n+m}{2}}_{12}$ 
with $N=n+m$ the total photon number (which is conserved, since we are considering a lossless BS). Clearly, we can only have the possibility for a  coincidence detection if $(n,m)$ are an (odd,odd) or (even,even) pair of integers, and so we will consider these two cases below.

\subsection{Multi-photons scattering diagram analysis for dual Fock state inputs}\label{sec:Scattering:Diagrams}
To build up intuition for how the multiphoton interference extended HOM effect comes about, and generalizes the 2-photon HOM effect, we consider in this section a series of input FS/FS states 
$\ket{n,m}_{12}$ to the beam splitter diagrammatically. Beginning with the well-known 2-photon HOM effect, we keep track of the number of transmitted and reflected photons from the input to the output modes. Since the beam splitter is a  2-port device, we can label separate component (partial) amplitudes to the total amplitude for the output coincidence state by indexing with the integer $k\in\{0,1,\ldots,n\}$. The index $k$  labels the number of photons \tit{transmitted} from input mode-1 to output mode-1, contributing the factor $t^k=(S_{11})^k$ to the $k$-th diagram. By considering dual Fock input states with larger and larger total photon number $N=n+m$, we will see a general pattern emerge, that will allow us in the subsequent section to analytically derive the extended HOM effect for an arbitrary dual Fock input state $\ket{n,m}_{12}$. For the choice of beam splitter scattering matrix we have chosen in \Eq{aout:intermsof:ain:line1} and \Eq{aout:intermsof:ain:line2}, the key matrix elements to consider are reflection coefficients $S_{12}=S_{21}=i\,r$, raised to either even or odd integer powers, for diagrams scattering different numbers of photons from mode-2 into mode-1, and mode-1 into mode-2, respectively.
We will pay particular attention to relative sign between ``mirage image" diagrams (described below) produced by different powers of the reflection coefficients $i\,r$ for such diagrams.

\subsubsection{The HOM effect: two-photon dual Fock  input $\ket{1,1}_{12}$; coincidence output $\ket{1,1}_{12}$}\label{sec:2:photon:input}
\begin{figure*}[th]
\begin{center}
\includegraphics[width=4.0in,height=1.75in]{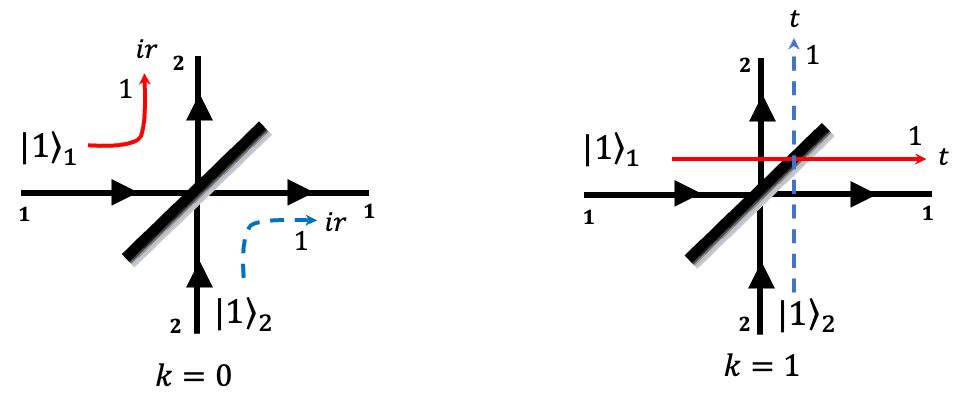}
%
%
\end{center} 
\caption{The 2-photon HOM effect with input $\ket{1,1}_{12}$ illustrating the two scattering amplitudes (left) $A_{k=0}$ both photons reflecting off the BS, and (right) $A_{k=1}$ both photons transmitting through the BS. 
Both amplitudes have equal magnitudes,  yet opposite signs, that combine to create destructive interference on the coincident output state $\ket{1,1}_{12}$. 
In general, the integer $k\in\{0,1,\ldots,n\}$ designates the number of  photons that transmit from input mode-1 to output mode-1 (with amplitude $t^k$)
for the dual Fock input state $\ket{n,m}_{12} = \ket{n}_1\,\ket{m}_2$. 
(Note: small (black) bold numbers on the axes indicate the input/output ports for mode-1 and mode-2. 
Small unbolded numbers indicate the number of photons transmitted and reflected by 
mode-1 (red curves) and mode-2 (blue curves)).
}
\label{fig:HOM:FS:FS:1:1}    
\end{figure*}

In \Fig{fig:HOM:FS:FS:1:1}  we consider the two-photon input state $\ket{1,1}_{12}$, 
apropos to the HOM effect \cite{HOM:1987}, and consider the 
total amplitude $A$ for coincidence on the output state $\ket{1,1}_{12}$. The total amplitude $A$ is composed of the sum of two components amplitudes $A = A_{k=0}+A_{k=1}$, where, in general, $k=\{0,\ldots,n\}$ (note: without loss of generality we will only consider the cases where $n\le m$) indicates the number of photons transmitted from input mode-1 to output mode-1 (giving rise to the contribution $t^k$ in $A_k$).  As is well known now, the HOM effect, i.e. the complete destructive interference of the quantum amplitudes for the output on the $\ket{1,1}_{12}$, is given by the sum of the contributions where (i) both photons are reflected into opposite numbered modes by the BS, with amplitude $A_{k=0}  = (i\,r)(i\,r)= -r^2$ as shown in \Fig{fig:HOM:FS:FS:1:1}(left), and both photons transmitted into their originating numbered modes, with amplitude $A_{k=1}  = t^2$, as shown in  \Fig{fig:HOM:FS:FS:1:1}(right). 
The crucial $(-1)$ sign in the $A_{k=0}$ amplitude comes from the contribution $(i\,r)^2$ resulting from the pair of reflecting photons, one from input mode-1 into output mode-2, and the other from input mode-2 to output mode-1.   
For a 50:50 beam splitter where $t=r=\tfrac{1}{\sqrt{2}}$
the two amplitudes $A_{k=0}$ and $A_{k=1}$ have equal magnitude, but opposite sign, and hence sum to zero:
\be{HOM:amplitude:sum}
A = A_{k=0} + A_{k=1} 
= \big[ (i\,r)^2 + t^2\big] 
= [-r^2 + t^2]\overset{t=r}{\to}0,
\ee 
(complete destructive interference)\cite{app:note:on:choice:of:S:in:HOM}. 
This is the famed HOM effect \cite{HOM:1987}. Of course, this is the ``textbook version" of the HOM effect, since we have implicitly assumed, that both photons 
(i) are completely indistinguishable (e.g. monochromatic in frequency, same spatial mode profile, etc\ldots), 
(ii) have arrived at the beam splitter at the same time (no relative time delay), and 
(iii) have been detected at the same time (no detection time difference). All these effects can be incorporated into the analysis of the HOM effect \cite{Legero_Rempe:2003}, and we will examine one of these considerations in \Sec{sec:1:beta:input}.

\subsubsection{The extended HOM effect: four-photon dual Fock input: input $\ket{1,3}_{12}$; coincidence output $\ket{2,2}_{12}$}\label{sec:4:photon:input:1:3:output:2:2}
\begin{figure*}[th]
\begin{center}
\includegraphics[width=4.5in,height=1.75in]{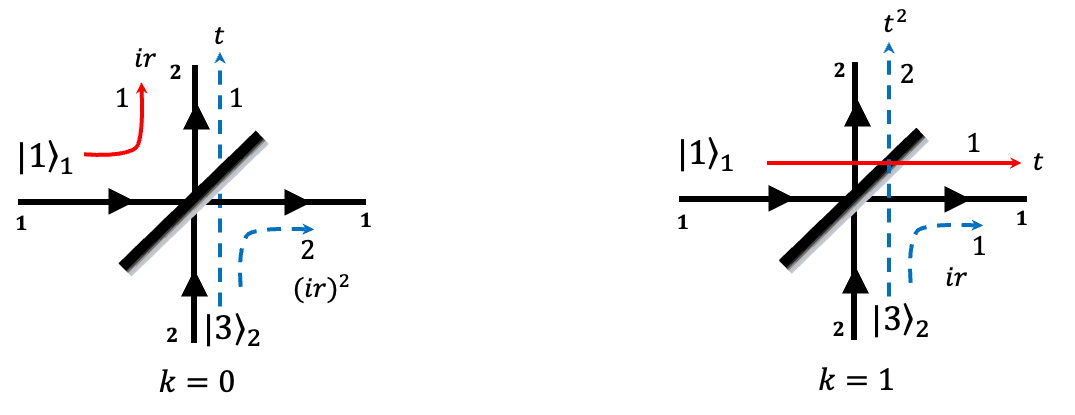}
%
%
\end{center} 
\caption{The 4-photon extended HOM effect with input $\ket{1,3}_{12}$ illustrating the two scattering amplitudes $A_{k=0}$ and $A_{k=1}$, with equal magnitude (when $t=r$) and opposite signs, that create destructive interference on the coincident output state $\ket{2,2}_{12}$. 
}
\label{fig:eHOM:FS:FS:1:3:in:2:2:out}    
\end{figure*}
In \Fig{fig:eHOM:FS:FS:1:3:in:2:2:out}   we consider the case of $N=4$ total number of photons input to the beam splitter, with state $\ket{n,m}_{12} = \ket{1,3}_{12}$. 
We note that \Fig{fig:eHOM:FS:FS:1:3:in:2:2:out}  with input state $\ket{1,3}_{12}$ is qualitatively similar to the HOM case in \Fig{fig:HOM:FS:FS:1:1} in that there are only two separate scattering diagrams, except now we must consider the scattering of $3$ photons from mode-2 (vs 1 photon scattering), and observe coincidence now on the output state $\ket{2,2}_{12}$. 
Tracking the scattering of $k$ photons from mode-1 (as in  \Fig{fig:HOM:FS:FS:1:1}), we again have two contributions to the amplitude, $A=A_{k=0}+A_{k=1}$.
Each component amplitude $A_k$ has the form of the products of all the matrix elements  
$\{S_{ij}\}$ of $S$, each raised to the power given by the number of photons 
that scatter from input mode-$j$ to output mode-$i$.

For $k=0$ \Fig{fig:eHOM:FS:FS:1:3:in:2:2:out}(left), one photon is reflected from mode-1 to mode-2, with scattering contribution $(S_{21})^1=i\,r$. Therefore, in order to have the possibility for coincidence on output state $\ket{2,2}_{12}$, one photon must be transmitted from mode-2 to mode-2, with contribution 
$(S_{22})^1=t$. The remaining 2  mode-2 input photons are then reflected into mode-1, with contribution 
$(S_{12})^2 = (i\,r)^2 = -r^2$. Since the photons are indistinguishable, there is no distinction between which of the 2 out of 3 mode-2 photons are reflected from mode-2 into mode-1. This gives rise to a combinatorial factor, indicated by the coefficient $C_0$, whose exact form is not of concern right now, except that it exists (we will return to its exact form in a \Sec{sec:General:Proof:eHOM}). Thus, from \Fig{fig:eHOM:FS:FS:1:3:in:2:2:out} (left), 
\bsub
\be{1:3:amplitude:sum:k:0}
A_{k=0} = C_0\,(S_{11})^{0}\,(S_{22})^{1}\,(S_{21})^{1}\,(S_{12})^{2}
=C_0\,(t)^0\,(t)^1\,(i\, r)^1\,(i\,r)^2 
= C_0\, t\,(i\,r)^3
\ee

For $A_{k=1}$ \Fig{fig:eHOM:FS:FS:1:3:in:2:2:out}(right), the opposite scattering scenario occurs, namely, the single mode-1 photon is transmitted into mode-1, with contribution $(S_{11})^1 = t$. 
Thus, in order to have coincidence,  one mode-2 photon must reflect into output mode-1, with scattering contribution  $(S_{12})^1 = i\,r$. The remaining two mode-2 photons are transmitted into mode-1 with contribution $(S_{22})^2 = (t)^2$. Thus, the contribution from this ``mirror image" scattering diagram, 
 \Fig{fig:eHOM:FS:FS:1:3:in:2:2:out} (right), is
\be{1:3:amplitude:sum:k:1}
A_{k=1}= (S_{11})^{1}\,(S_{22})^{2}\,(S_{21})^{0}\,(S_{12})^{1}  
=C_1\,(t)^1\,(t)^2\,(i\, r)^0\,(i\,r)^1
= C_0 t^3\,(i\,r),
\ee 
where by symmetry (as we shall prove later) we have $C_0=C_1$.
\Fig{fig:eHOM:FS:FS:1:3:in:2:2:out}(right) is the ``mirror image" of the diagram \Fig{fig:eHOM:FS:FS:1:3:in:2:2:out}(left) in the sense that the number of photons transmitted/reflected  is swapped into the number reflected/transmitted respectively, for each of mode-1 and mode-2. 
We also note that $A_{k=1}$ is obtained from $A_{k=0}$ by swapping $t\leftrightarrow (i\,r)$, and invoking the combinatorial symmetry $C_k=C_{n-k}$.
These mirror image diagrams will be crucial for understanding the extended HOM effect, since they generalize the two mirror image diagrams in \Fig{fig:HOM:FS:FS:1:1} for the HOM effect. 

By symmetry (as well shall prove later) of \Fig{fig:eHOM:FS:FS:1:3:in:2:2:out} we have $C_0=C_1$, so that the final amplitude is 
\be{1:3:amplitude:sum:total}
A = C_0 \big[t\,(i\,r)^3 + t^3\,(i\,r)\big]
= C_0\,t\,(i\,r)\, \big[-r^2 + t^2\big]
\overset{t=r}{\to} 0,
\ee 
\esub
where the latter follows for a 50:50 beam splitter where $t=r=\tfrac{1}{\sqrt{2}}$, and $i^2=-1$. Thus, as in the HOM effect, the two amplitude contributions cancel due to a relative sign change arising from the different number of photons that reflect into opposite modes (input mode-1 to output mode-2, and input mode-2 to output mode-1).

Note that in first equalities in \Eq{1:3:amplitude:sum:k:0} and \Eq{1:3:amplitude:sum:k:1} we have made use of the matrix elements $S_{ij}$ as a bookkeeping device, mainly to keep track of their exponents indicating the number of photons scattering (transmitting or reflecting) from input mode-$j$ into output mode-$i$. However, it is easy enough to write down the partial  amplitude $A_{k}$ directly (the last equalities in \Eq{1:3:amplitude:sum:k:0} and \Eq{1:3:amplitude:sum:k:1}) by simply raising $t$ to the combined number of photons transmitted in the diagram, and raising $(i\,r)$ to the combined number of photons reflected in the same diagram. 
In the future, we will dispense with the $S_{ij}$ notation, and write down the component amplitude $A_k$ directly, by this later observation.
Note that for each diagram $A_k$, the photons can be considered as indistinguishable scattering ``billiard balls" (hence the presence of a combinatorial factor). The quantum nature of interference manifests itself in the addition of the component amplitudes before squaring, in order to obtain a probability.

\subsubsection{The extended HOM effect: four-photon dual Fock input: input $\ket{2,2}_{12}$; coincidence output $\ket{2,2}_{12}$}\label{sec:4:photon:input2:2:output:2:2}
\begin{figure*}[th]
\begin{center}
\includegraphics[width=5.5in,height=1.75in]{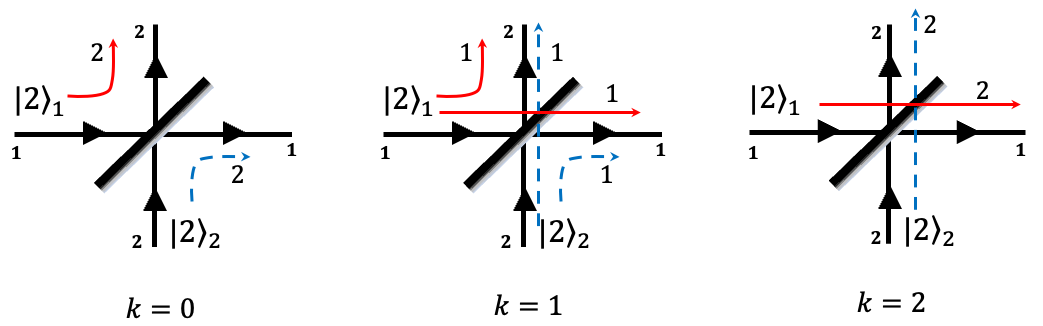}
%
\end{center} 
\caption{The 4-photon extended HOM effect with input $\ket{2,2}_{12}$ illustrating the three scattering amplitudes, with outer two amplitudes $A_{k=0}$ and $A_{k=2}$ of equal magnitude (when $t=r$)
same sign, and ``unpaired" central amplitude $A_{k=1}$, leading to 
that no destructive interference on the coincident output state $\ket{2,2}_{12}$. 
}
\label{fig:eHOM:FS:FS:2:2:in:2:2:out}    
\end{figure*}

\Fig{fig:eHOM:FS:FS:2:2:in:2:2:out} is again a case with $N=4$ total number of photon, similar to 
\Fig{fig:eHOM:FS:FS:1:3:in:2:2:out}, except now the input state is  $\ket{2,2}_{12}$ (with the same output coincidence state $\ket{2,2}_{12}$). For this case $\ket{n,m}_{12}$, with $(n,m)$ (even,even), we now crucially have an \tit{odd} number partial scattering contributions to the total amplitude, namely
$A = A_{k=0}+A_{k=1}+A_{k=2}$ (again, $k$ counts the scattering (transmission) of $k$ mode-1 photons into mode-1). This time, the ``mirror image" scattering contributions 
$A_{k=0} = C_0 \,(i\,r)^4= C_0\,r^4$ and $A_{k=2} = C_2\, t^4$ (where by symmetry $C_0=C_2$, as shown later) 
have the \tit{same} sign, and hence \tit{add} (constructive interference), rather and cancel each other. The middle diagram $A_{k=1}$ in \Fig{fig:eHOM:FS:FS:2:2:in:2:2:out} now contains the relative $i^2=(-1)$ sign change, due to the pair of photons scattering from input mode-1 to output mode-2, and input mode-1 to output mode-2 with contribution 
$(i\,r)^2=-r^2$. However, since this middle diagram in ``unpaired" with any other diagram (since there are no other diagrams remaining, it's contribution $A_{k=1}= C_1\, t^2\,(i\,r)^2 = -C_1\,t^2\,r^2$  remains ``un-canceled." Thus, the total amplitude is given by 
\be{2:2:amplitude:sum:total}
A = (A_{k=0}+A_{k=2}) + A_{k=1} 
= C_0\,(r^4 + t^4)+C_1(-t^2\,r^2) 
\overset{t=r}{\to} (2C_0-C_1)\,t^2\,r^2 \ne0, 
\ee
since $C_1\ne C_0=C_2$ (shown later).

\subsubsection{The extended HOM effect: eight-photon dual Fock input: input $\ket{3,5}_{12}$; coincidence output $\ket{4,4}_{12}$}\label{sec:4:photon:input:3:5:output:4:4}
The important point of note of the previous two subsections is the following. When $\ket{n,m}_{12}$ has $(n,m)$ (even, even), there are (i) an odd number of scattering diagram contributions $A_k$ to the over scattering amplitude 
$A=\sum_{k=0}^{n (even)} A_k$, (ii) 
matching ``mirror-image" scattering diagrams have the \tit{same} sign (vs opposite signs) and hence do not cancel, and (iii) there will always be an ``un-paired" middle diagram $A_{k=(n+m)/2-1}$ which will not be able to cancel with any other diagram.  

On the other hand,  when $\ket{n,m}_{12}$ has $(n,m)$ (odd,odd), ``mirror-image"   amplitude diagrams 
have the same magnitudes, yet \tit{opposite} signs, and hence \tit{cancel pairwise}.

\begin{figure*}[th]
\begin{center}
\hspace{-0.55in}
\includegraphics[width=7.0in,height=1.5in]{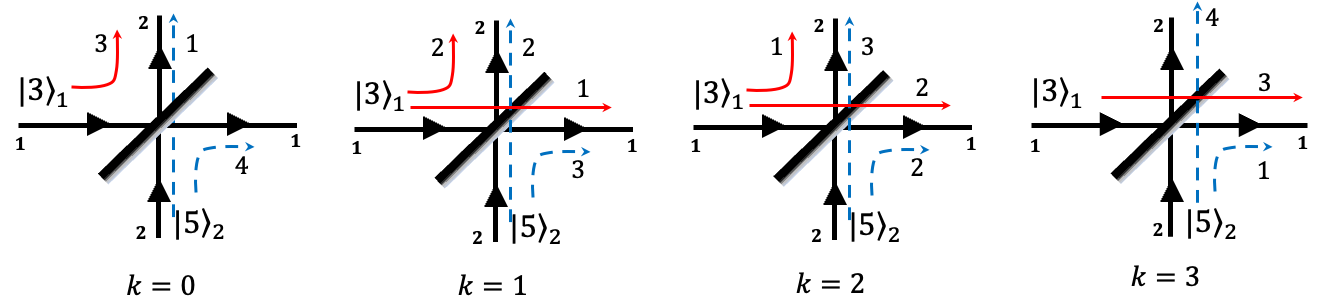}
%
%
\end{center} 
\caption{The 8-photon extended HOM effect with input $\ket{3,5}_{12}$ illustrating the two pairs of scattering amplitudes, $(A_{k=0}, A_{k=3})$, and $(A_{k=1}, A_{k=2})$, 
each with with equal magnitude and opposite signs (when $t=r$), that cancel in pairs, and contribute to the complete destructive interference on the coincident output state $\ket{4,4}_{12}$. 
}
\label{fig:eHOM:FS:FS:3:5:in:4:4:out}    
\end{figure*}

To verify this intuition, before turning to the general case, we consider the higher order multi-photon case in \Fig{fig:eHOM:FS:FS:3:5:in:4:4:out} with (odd, odd) dual Fock input $\ket{3,5}_{12}$, 
with output coincidence state $\ket{4,4}_{12}$. The important point of this figure, is that the total amplitude $A$ on the coincidence state groups itself into two pairs: 
\be{3:5:amplitude:sum:formal}
A=(A_{k=0} + A_{k=3}) + (A_{k=1} + A_{k=2}),
\ee with each pair $(A_{k} + A_{n-k})$ for $k=0,\ldots,\half(n-1)$ \tit{canceling separately}. 

The amplitude component $A_{k=0}$ \Fig{fig:eHOM:FS:FS:3:5:in:4:4:out}(leftmost) has zero mode-1 photons transmitted into mode-1, and all three mode-1 photons reflecting into mode-2 with contribution $(i\,r)^3$. 
This implies that four mode-2 photons must reflect into mode-1 with contribution 
$ (i\,r)^4$. The total amplitude contribution for this diagram is then   $A_{k=0}=C_0\, t\,(i\,r)^7$. 
It's ``mirror image"  $A_{k=3}$ \Fig{fig:eHOM:FS:FS:3:5:in:4:4:out}(rightmost)  has all three mode-1 photons transmitting into mode-1, while none are reflected into mode-2. This implies that one mode-2 photon reflects into mode-1 with contribution $i\,r$.
Thus, the total amplitude contribution for this diagram is  $A_{k=3}=C_3\,t^7\,(i\,r)$. 
Again, symmetry dictates that $C_{k} = C_{n-k}$ and hence these two terms cancel pairwise for a 50:50 beam splitter when $t=r$, 
\be{3:5:amplitude:sum:k:0:k:3}
A_{k=0}+A_{k=3} 
= C_0\, \big[ t\,(i\,r)^7+t^7\,(i\,r) \big] 
= C_0\,t\,(i\,r)\, \big[-r^6+\,t^6 \big]\overset{t=r}{=} 0,
\ee
using $i^6=i^2=-1$ in the last equality.

Examining the pair $A_{k=1} + A_{k=2}$ in \Fig{fig:eHOM:FS:FS:3:5:in:4:4:out}(middle left, middle right), we find a similar pair cancelation of the diagrams  $A_{k=1}$ for one mode-1 photon transmitting into mode-1 (and two photons reflecting into mode-2), and its mirror image amplitude diagram $A_{k=2}$. 
By multiplying $t$ and $i\,r$ by total number of transmitted and reflected photons, respectively, we see from 
 \Fig{fig:eHOM:FS:FS:3:5:in:4:4:out} that the $k=1$ diagram contributes an amplitude
 $A_{k=1} = C_1\,t^3\,(i\,r)^5$, while the $k=2$ diagram contributes an amplitude
 $A_{k=2} = C_2\,t^5\,(i\,r)^3$ (with $C_2=C_1$, again by symmetry).
The sum of the two diagrams  yields the overall contribution 
\be{3:5:amplitude:sum:k:1:k:2}
A_{k=1}+A_{k=2}
=C_1\,\big[t^3\,(i\,r)^5 + t^5\,(i\,r)^3\big]
=C_1\,t^3\,(i\,r)^3\,\big[-r^2 + t^2\big]
\overset{t=r}{\to}0, 
\ee
using $i^2=-1$ in the last equality.
In a sense, it could be said that in the extended HOM effect,  the HOM effect (a cancelation of two mirror-image diagrams) occurs on multiple pairs of component amplitudes, simultaneously.

\subsubsection{The extended HOM effect: general odd-odd dual Fock input: $\ket{2n+1,2m+1}_{12}$; coincidence output $\ket{n+m+1,n+m+1}_{12}$}\label{sec:4:photon:input:3:5:output:4:4}
\begin{figure*}[!ht]
\begin{center}
\includegraphics[width=5.5in,height=1.75in]{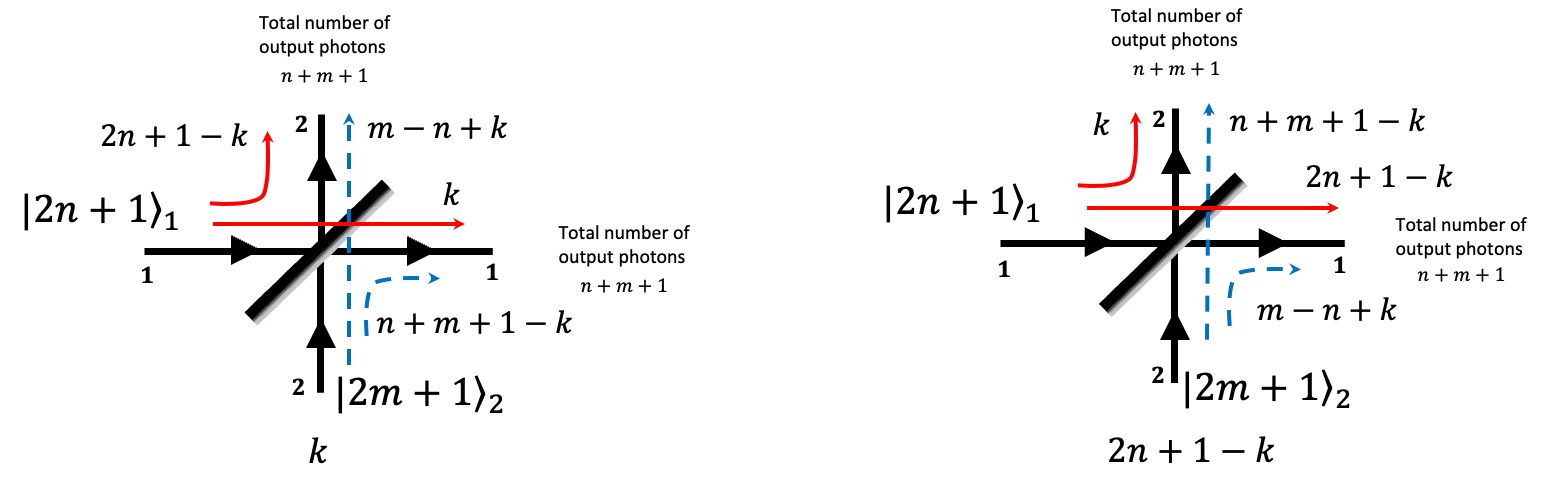}
\end{center} 
\caption{The odd-odd-photon extended HOM effect with input $\ket{2n+1,2m+1}_{12}$ illustrating the two ``outer" scattering amplitudes $A_{k}$ and $A_{2n+1-k}$, with equal magnitude and opposite signs (when $t=r$) that cancel each other, contributing to the  complete destructive interference on the coincident output state $\ket{n+m+1,n+m+1}_{12}$.
}
\label{fig:eHOM:FS:FS:odd:odd:in:k:2np1mk:ampls}    
\end{figure*}

To show that \Fig{fig:eHOM:FS:FS:3:5:in:4:4:out}  is indeed the general trend of multiple pairwise amplitude cancellations for mirror image component amplitude diagrams for odd/odd dual Fock inputs 
we consider in 
\Fig{fig:eHOM:FS:FS:odd:odd:in:k:2np1mk:ampls}  with the general odd-odd dual Fock input 
$\ket{2n+1,2m+1}_{12}$, with output coincident state $\ket{n+m+1,n+m+1}_{12}$. 
We consider the amplitude $A_{k}$ \Fig{fig:eHOM:FS:FS:odd:odd:in:k:2np1mk:ampls}(left, red curves)  with (by definition) $k\in\{0,1,\ldots,2n+1\}$ photons transmitted from input mode-1 to output mode-1, and hence $2n+1-k$ photons reflected from input mode-1 into output mode-2. 
It's mirror image amplitude diagram \Fig{fig:eHOM:FS:FS:odd:odd:in:k:2np1mk:ampls}(right, red curves)
is $A_{2n+1-k}$ swaps the number of photons transmitted with those reflected 
thus having  $2n+1-k$ photons transmitted from input mode-1 to output mode-1, and hence $k$ photons reflected from input mode-1 into output mode-2.

In order to have the output coincidence state $\ket{n+m+1,n+m+1}_{12}$ for amplitude $A_{k}$, 
$m-n+k$ photons must be transmitted from input mode-2 to output mode-2, and $n+m+1-k$ 
photons must be reflected from input mode-2 to output mode-1, as shown in 
\Fig{fig:eHOM:FS:FS:odd:odd:in:k:2np1mk:ampls}(left, blue curves).
Multiplying $t$ by the total number of photons transmitted, and $i\,r$ by the total number of photons reflected we have $A_{k} = C_k\, (t)^{m-n+2k}\,(i\,r)^{3n+m+2-2k}$. 

For the mirror image 
amplitude $A_{2n+1-k}$ \Fig{fig:eHOM:FS:FS:odd:odd:in:k:2np1mk:ampls}(right), the contribution is simply given by that of  $A_{k} $ but with $t\leftrightarrow (i\,r)$, yielding $A_{2n+1-k} = C_k\, (i\,r)^{m-n+2k}\,(t)^{3n+m+2-2k}$ 
(again with $C_{2n+1-k} = C_k$ by symmetry). Thus, thus sum of the two amplitudes is given by
\bea{Ak:plus:A2np1mk}
A_{k} + A_{2n+1-k} &=& C_k\,(t)^{m-n+2k}\,(i\,r)^{m-n+2k}
\big[
(i\,r)^{2+4(n-k)} + (t)^{2+4(n-k)}
\big], \no
&=&
C_k\,(t)^{m-n+2k}\,(i\,r)^{m-n+2k}
\big[
-(r)^{2+4(n-k)} + (t)^{2+4(n-k)}
\big]\overset{t=r}{\to} 0,
\eea
where the crucial relative minus sign in the last line of \Eq{Ak:plus:A2np1mk}
 arises from $i^{2+4(n-k)} = (i^2)\,(i^4)^{n-k}=(-1)(1)=-1$.
\Eq{Ak:plus:A2np1mk} holds for all $k\in\{0,1,\ldots,2n+1\}$, and demonstrates that for odd/odd dual Fock inputs, the extended HOM destructive interference effect on the output coincident state proceeds by pairwise HOM-like cancellations of amplitudes on mirror image diagrams.  These mirror image diagrams swap the number of photons transmitted/reflected with the number 
 reflected/transmitted respectively, for both modes 1 and 2.

\subsection{General analytical proof of the extended HOM for a 50:50 beam splitter}\label{sec:General:Proof:eHOM}
Armed with the insights from the analysis of the multi-photon scattering amplitude diagrams  in the previous 
\Sec{sec:Scattering:Diagrams},  we now construct a general proof of the extended HOM effect for both (odd,odd) and (even,even) dual Fock inputs to lossless 50:50 BS.


As a two-port device with modes 1 and 2 associated with the photon creation operators  $a^\dag_1$ and $a^\dag_2$, the 50:50 beam splitter (with $t=r = \tfrac{1}{\sqrt{2}}$) enacts the transformation 
(now writing input modes in terms of output modes, 
using the inverse of \Eq{aout:intermsof:ain:line1} and \Eq{aout:intermsof:ain:line2})
\bsub
\bea{ain:intermsof:aout}
a^\dag_{1,in} &=& \tfrac{1}{\sqrt{2}}\,(a^\dag_{1,out} - i\, a^\dag_{2,out}), \\
a^\dag_{2,in} &=& \tfrac{1}{\sqrt{2}}\,(a^\dag_{2,out} - i\, a^\dag_{1,out}).
\eea
\esub
Consider the dual Fock input state $\ket{n,m}^{(in)}_{12}= 
\frac{(a^\dag_{1,in})^n}{\sqrt{n!}}\, \frac{(a^\dag_{2,in})^m}{\sqrt{m!}}\,\ket{0,0}_{12}$
for total photon number $N=n+m$.
Without loss of generality, we consider the case $n\le m$ (the derivation is similar for $n\ge m$).
The output state $\ket{n,m}^{(out)}_{12}$ for a 50:50 beam splitter is then simply given by
replacing the input creation operators in terms of the output creation operators as given above (and then for notational convenience, dropping the $out$ subscript) to yield
\bsub
\bea{n:m:out}
\ket{n,m}^{(out)}_{12} &=& \,\dfrac{(a_1^\dag - i\, a_2^\dag)^n}{\sqrt{2^n\, n!}}\,
                                             \dfrac{(a_2^\dag - i\, a_1^\dag)^m}{\sqrt{2^m\,m!}} \, \ket{0,0}_{12}, \no 
&=& \frac{1}{\sqrt{2^{n+m}\,n!\,m!}}\,
        \sum_{k=0}^{n} \binom{n}{k}\,(a_1^\dag)^{k}\,(-i\,a_2^\dag)^{n-k}\,
        \sum_{\l=0}^{m} \binom{m}{\l}\,(-i\,a_1^\dag)^{\l}\,(a_2^\dag)^{m-\l}\, \ket{0,0}_{12}, \no 
 &=& \frac{(-i)^{n}}{\sqrt{2^{n+m}\,n!\,m!}}\,
         \sum_{k=0}^{n}  \sum_{\l=0}^{m}\,\binom{n}{k}\, \binom{m}{\l}\,(-i)^{\l-k}\,
 				  (a_1^\dag)^{k+\l} \,(a_2^\dag)^{n+m-(k+\l)}\, \ket{0,0}_{12}.
\eea
\esub
We are interested in the coincidence output state component $\ket{\frac{n+m}{2},\frac{n+m}{2}}_{12}$ (i.e. equal number of photons in the output modes $a$ and $b$).
This implies that $n$ and $m$ are either both odd, or both even, otherwise $\frac{n+m}{2}$ is a half integer and there is trivially no possibility for a coincidence state in the beam splitter output. 
Examining \Eq{n:m:out}, we note that the component $\mathcal{C}_{(n+m)/2,(n+m)/2}\,\ket{\frac{n+m}{2},\frac{n+m}{2}}_{12}$ of the full output coincident state $\ket{n,m}^{(out)}_{12}$ in \Eq{n:m:out}
is given the condition that $k+\l = (n+m)/2$.
Thus, we have:
\bsub
\bea{S:general}
\hspace{-0.5in}
\mathcal{C}_{(n+m)/2,(n+m)/2}\, \ket{\thalf(n+m),\thalf(n+m)}_{12}
&=& 
S'\, 
\frac{(-i)^{n}\,[\thalf(n+m)]!}{\sqrt{2^{n+m}\,n!\,m!}}\,
\frac{(a_1^\dag)^{\frac{n+m}{2}}}{\sqrt{(\frac{n+m}{2})!}}\,
\frac{(a_2^\dag)^{\frac{n+m}{2}}}{\sqrt{(\frac{n+m}{2})!}}\, \ket{0,0}_{12}, \\ 
S' &\defn&  \sum_{k=0}^{n}  \sum_{\l=0}^{m} \delta_{k+\l,\frac{n+m}{2}}\,
                                                                        \binom{n}{k}\, \binom{m}{\l}\,(-i)^{\l-k} 
                                                                        \equiv (-i)^{n+\tfrac{n+m}{2}}\,S,\qquad\qquad \label{S:general:2}\\
S &\defn&    \sum_{k=0}^{n} \binom{n}{k}\,\binom{m}{\frac{n+m}{2}-k}\, (-1)^{k}, \label{S:general:3}
\eea
\esub
where the use of $ \delta_{k+\l,\frac{n+m}{2}}$ in  \Eq{S:general:2} implies
$(-i)^{\l-k}\to(-i)^{(n+m)/2}\,(-i)^{-2k}$, and we note that $(-i)^{-2k} = (i^2)^k = (-1)^k$ which crucially appears in the summand in \Eq{S:general:3}. 

In \Eq{S:general:3} have defined the sum $S$ as the total scattering amplitude $A$ 
for coincidence detection, as in the previous sections 
(modulo unimportant overall factors of $-i$). 
(Note: in general the upper limit of $k$ in $S$ is given by $\trm{min}(n, \frac{n+m}{2}) = n$ 
(from the ranges of $k$ in the denominators of the binomial coefficients), since
$\frac{n+m}{2}-n = \frac{m-n}{2}\ge 0$ by assumption that $m\ge n$).
Each term in the sum $S$ corresponds to the amplitude diagram contribution $A_k$ in the previous section where we have already set $t=r=\tfrac{1}{\sqrt{2}}$. The combinatorial factors $C_k$ discussed in the previous sections are given by the product of the binomial coefficients (one for each mode) appearing in $S$ in \Eq{S:general:3}.

We now examine two cases for $n$ even or $n$ odd. The lesson learned from the previous sections is that the total  $S=A$ amplitude for the output coincident state should be grouped together with each diagram $A_k$ paired with its mirror image diagram $A_{n-k}$, which then cancel pairwise for a 50:50 beam splitter if $n$ is odd, yet add constructively if $n$ is even.

\subsubsection{\tbf{$\bf{n}$ odd:}}
Since $n$ is odd, there are and even number of terms in the sum $S$ (indexed by $k\in\{0,\ldots,n\}$. Let us break $S$ into two sums $S_1+S_2$ with an equal number of terms in each sum, 
$k\in\{0,\ldots,\frac{n-1}{2}\}\cup\{\frac{n+1}{2},\ldots,n\}$ (e.g. for $n=7$, 
$k\in \{0,1,2,3\}\cup\{4,5,6,7\}$)
\be{S1:S2}
S = S_1 + S_2 = \sum_{k=0}^{\frac{n-1}{2}} \binom{n}{k}\,\binom{m}{\frac{n+m}{2}-k}\, (-1)^{k}  
                           +    \sum_{k=\frac{n+1}{2}}^{n} \binom{n}{k}\,\binom{m}{\frac{n+m}{2}-k}\, (-1)^{k}.
\ee
In the second sum $S_2$ let $k'=n-k$ so that $k=n-k'$, which yields
\bsub
\bea{S2}
S_2 &=&  \sum_{k'=\frac{n-1}{2}}^{0} \binom{n}{n-k'}\,\binom{m}{\frac{n+m}{2}-n+k'}\, (-1)^{n-k'},  \label{S2:line:1} \\
&=&(-1)^n\, \sum_{k'=0}^{\frac{n-1}{2}} \binom{n}{k'}\,\binom{m}{\frac{n+m}{2}-k'}\, (-1)^{n-k'}, \label{S2:line:2} \\
&\equiv& (-1)^n\, S_1, \label{S2:line:3}
\eea
\esub
where in going from \Eq{S2:line:1} to the \Eq{S2:line:2}  we have used that 
$\binom{n}{k'} = \binom{n}{n-k'}$ and
$\binom{m}{\frac{n+m}{2}-n+k'} = \binom{m}{m-(\frac{n+m}{2}-n+k')}=\binom{m}{\frac{n+m}{2}-k'}$, 
(which proves the claim in the previous sections that $C_k = C_{n-k}$),
and  $(-1)^{-k'} = (-1)^{k'}$.
The crucial result is that since we have assumed that $n$ is \tit{odd}, the multiplicative factor in \Eq{S2:line:3} is $(-1)^n = -1$, so that $S_2 = -S_1$, and hence $S=S_1+S_2 = 0$. This is the extended HOM effect\cite{eHOM_PRA:2022}.

Note that the sum $S$ in \Eq{S:general:3} cancels pairwise 
due to the alternating sign $(-1)^k$ canceling those pairs with equal magnitudes (the product of the binomial coefficients in \Eq{S:general:3}). This is exactly the canceling of the amplitude $A_k$ with it's mirror image amplitude $A_{n-k}$, demonstrated diagramaticaly in the previous section. 
Put another way, for $n$ odd, $S=\sum_{k=0}^{(n-1)/2} (A_k + A_{n-k})$, with each pair of terms $(A_k + A_{n-k})$ in the summand canceling separately.

\subsubsection{\tbf{$\bf{n}$ even:}}
In the case of $n$ even, $S$ now contains an odd number of terms indexed by $k\in\{0,\ldots,n\}$ which we now break up into $S=S_1 + T_{n/2} + S_2$ via
$k\in\{0,\ldots,\frac{n}{2}-1\}\cup \{\frac{n}{2}\}\cup \{\frac{n}{2}+1,\ldots,n\}$ 
(e.g. for $n=8$, $k\in \{0,1,2,3\}\cup \{4\}\cup\{5,6,7,8\}$).
A calculation exactly analogous to the previous case yields
\bsub
\bea{S:n:even}
S &=& S_1 + (-1)^{n/2}\,\binom{n}{n/2}\,\binom{m}{m/2} + (-1)^n\,S_1, \label{S:n:even:line:1} \\
&=& 2 S_1 + (-1)^{n/2}\,\binom{n}{n/2}\,\binom{m}{m/2}, 
\qquad S_1 = \sum_{k=0}^{\frac{n}{2}-1} \binom{n}{k}\,\binom{m}{\frac{n+m}{2}-k}\, (-1)^{k}, \label{S:n:even:line:2}
\eea
\esub
where we have used the crucial fact that since we have assumed that $n=2n'$ is even, then $(-1)^n=1$ in \Eq{S:n:even:line:1}.
Thus, even if $S_1$ were to sum to zero (which could only possibly occur for $n'-1$ odd, or equivalently $n'$ even), one would still have $|S|^2\ge|T_{n/2}|^2>0$. 
Thus, in general, for $n$ even we have 
$|S|^2>0$ \cite{eHOM_PRA:2022}, and there is not complete destructive interference on the output component state $\ket{\frac{n+m}{2},\frac{n+m}{2}}_{12}$.

An implication of the above results is that a measurement of output component state $\ket{2,2}_{12}$ for the input state $\ket{1,3}^{(in)}_{12}$ will exhibit complete destructive interference. However, a measurement of output component state $\ket{2,2}_{12}$ for the input state $\ket{2,2}^{(in)}_{12}$ will not exhibit complete destructive interference, as shown in the previous sections. From the above results on dual Fock inputs, one deduces  that if the mode-1 input state is of odd parity (i.e. composed of only odd number of photons), then regardless of the mode-2 input state, pure or mixed, there will be a central nodal line of complete destructive interference (zeros) along the diagonal of the output joint photon number probability distribution of a balanced 50:50 BS. This is the statement of the extended HOM effect\cite{eHOM_PRA:2022}.

%
 %
 \section{The role of imperfect detection efficiency: Fock state/Coherent State input to a 50:50 beam splitter}\label{sec:1:beta:input}
In a realistic experiment one has to contend with the prospects of imperfect detector efficiency, the influence of the photon mode functions in wavepackets, and potential time delay between photon detection in the output ports of the BS. In this section we consider two of these aspects in the context of a possible experimental realization of the extended HOM effect for the case of a Fock/coherent state input $\ket{1,\beta}_{12}$ with a single  photon $\ket{1}_1$ entering port-1 of a 50:50 BS, and a coherent state  
$\ket{\beta}_2  = e^{-|\beta|^2/2} \,\sum_{m=0}^\infty \tfrac{(\beta)^m}{\sqrt{m!}}\,\ket{m}_2$ of (complex) amplitude $\beta$, with mean photon number $\bar{n}_2 = |\beta|^2\,$
\cite{Scully_Zubairy:1997,Loudon:2000,Agarwal:2013,Ou_Book:2017,Gerry_Knight:2023}.  
(The effects of the mode function wavepackets, and time delays between detected photons are explored in a separate forthcoming publication \cite{eHOM:Is:Odd:Full:In:Prep:2024}).
The coherent state represents an idealized, continuous wave (CW) laser of fixed frequency $\omega_2$, and has the property that it is an eigenstate of the mode-2 annihilation operator, $a_2\,\ket{\beta}_2 = \beta\,\ket{\beta}_2$. 
The single photon Fock state could be generated for example by the heralding on one component ($1'$) of a weak two-mode squeezed state
$\ket{TMSS}_{11'} = \frac{1}{\cosh(r)}\sum_{n=0}^\infty \tanh^n(r)\ket{n,n}_{11'}$ where $r$ is the squeezing parameter. This state can be generated for example, by the nonlinear processes of spontaneous parametric down-conversion or by four-wave mixing \cite{Boyd:1991,Scully_Zubairy:1997,Agarwal:2013,Ou_Book:2017,Gerry_Knight:2023,Boyd:1991}.
For a weak field two-mode squeezed state, a detection in mode-$1'$ ``heralds" the correlated presence of a single photon in mode-1, which can then be fed forward into the mode-1 input port of a 50:50 BS.

The probability $P(N_1,N_2)$ to perfectly detect $N_1$ photons from mode-1 and $N_2$ photons from mode-2 from the output of a lossless 50:50 beam splitter is given by $P(N_1,N_2) = |\A_{N_1,N_2}|^2$, where the quantum amplitude is given by $\A_{N_1,N_2} = {}_{12}\IP{N_1,N_2}{1,\beta}^{out}_{12}$. For the input state 
\be{1:beta:input:state}
\ket{1,\beta}^{in}_{12} = a^\dag_1\,D_2(\beta)\ket{0,0}_{12},
\ee
 where 
$D_2(\beta) = e^{\beta\,a^\dag_2 - \beta^*\,a_2}$ is the mode-2 displacement operator, defined such that its action on the vacuum state produces a CS,
$D_2(\beta)\ket{0}_2=\ket{\beta}_2$ \cite{Scully_Zubairy:1997,Agarwal:2013,Ou_Book:2017,Gerry_Knight:2023}. As in 
\Sec{sec:General:Proof:eHOM}, the action of the beam splitter is affected by writing the input creation operators in terms of the output creation operators (and again, dropping $in$ and $out$ subscripts for clarity) producing 
\be{1:beta:output:state}
\ket{1,\beta}^{out}_{12} = \frac{1}{\sqrt{2}} (a^\dag_1+a^\dag_2)\,D_1(\tfrac{-\beta}{\sqrt{2}})\,D_2(\tfrac{\beta}{\sqrt{2}})\, \ket{0,0}_{12} =\frac{1}{\sqrt{2}} (a^\dag_1+a^\dag_2), \ket{\tfrac{-\beta}{\sqrt{2}},\tfrac{\beta}{\sqrt{2}}}_{12}, 
\ee
where the transformation of $D_2(\beta)$ by the BS, and the independence of mode-1 and mode-2 operators, creates the (tensor) product of displacement of operators
$D_1(\tfrac{-\beta}{\sqrt{2}})\,D_2(\tfrac{\beta}{\sqrt{2}})$, leading to a product of mode-1/mode-2 coherent states, each with reduced amplitudes $\tfrac{-\beta}{\sqrt{2}}$ and $\tfrac{\beta}{\sqrt{2}}$, respectively. Using the Hermitian conjugate of the boson operator relations that $a_i\ket{N}_i = \sqrt{N}\,\ket{N-1}_i$, namely
${}_{i}\bra{N}\, a^\dag_i = \sqrt{N}\,{}_{i}\bra{N-1}$, it is easy to show that 
\be{P:N1:N2}
P_{12}(N_1,N_2) = |{}_{12}\IP{N_1,N_2}{1,\beta}^{out}_{12}|^2
= \frac{e^{-|\beta|^2}}{N_1!\,N_2!\,2^{N_1+N_2}}\,(N_1-N_2)^2,   
\ee
which is a straightforward extension \cite{Ou:1996:Eq:4:note} of a  result first shown by Ou in 1996 \cite{Ou:1996} 
for the input state $\ket{1,N}_{12}$. 
This result also appears as Eq.(8) in the 2012 work Birrittella, Mimih and Gerry\cite{BMG:2012,Gerry_Knight:2023}.
The salient point here is that for $N_1=N_2$ one has a central nodal line of zeros along the main diagonal of the joint output probability distribution $P_{12}(N_1,N_2)$, which is 
shown in \Fig{fig:eHOM:fig2:fig3:left:column:1photon:CS} (top row, middle) for $n=1$.

To model imperfect detection efficiency \cite{Loudon:1983,Scully_Zubairy:1997, Knight:2002,eHOM_PRA:2022}
consider the detection of $n$ photons in the output of a single mode, with detector efficiency $0\le\eta\le1$. 
The relevant point is that these $n$ detected photons could have resulted from $N\ge n$ photons impinging on the detector, of which only $n$ were actually registered, due the finite efficiency of the detector. 
The the probability that $n$ photons were detected, which $N-n$ were not, is the Bernoulli factor
$\binom{N}{n}\,\eta^n\,(1-\eta)^{N-n}$, where the binomial coefficient indicates the indistinguishability of which $n$ of the total of the impinging $N$ photons were actually detected. The total probability $P_\eta(n)$ is then a sum over all possible values of $N\ge n$, namely
\be{P:eta} 
P_\eta(n) = \sum_{N=n}^\infty\binom{N}{n}\,\eta^n\,(1-\eta)^{N-n}\,P_N,
\ee
 where 
$P_N=|\mathcal{A}_N|^2$ is the probability to perfectly detect $N$ photons 
\cite{quantum:efficiency:note}. 
%
Applying this to each output mode of the BS, and assuming equal detector efficiencies, the joint probability $P_\eta(n,n)$
to measure $n$ coincidence counts from the output of the beam splitter for the input $\ket{1,\beta}_{12}$ is given by
\bea{P:eta:n:n}
P_\eta(n,n) &=&\sum_{N_1=n}^\infty\, \sum_{N_2=n}^\infty\,\binom{N_1}{n}\,\binom{N_2}{n}\,
\eta^{2\,n}\,(1-\eta)^{N_1+N_2-2n}\,P_{12}(N_1,N_2), \no
&=& 2\,\eta^{2\,n}\sum_{N_1>N_2\ge n}^\infty\,\binom{N_1}{n}\,\binom{N_2}{n}\,
\eta^{2\,n}\,(1-\eta)^{N_1+N_2-2n}\,P_{12}(N_1,N_2)> 0.
\eea
In the last line of \Eq{P:eta:n:n} we have broken the double sum into two pieces: (i) a single diagonal sum over $N_1=N_2$, which is zero, since $P_{12}(N_1,N_1)=0$ by \Eq{P:N1:N2}, and (ii) the remaining off-diagonal double sum, now with $N_1>N_2\ge n$, with a factor of $2$ out front (due to the symmetry of the entire expression under the exchange $N_1\leftrightarrow N_2$). Thus, while the presence of the CNL can be observed, it is no longer exactly zero. In addition, detecting the coincidence output state $\ket{n,n}_{12}$ is proportional to $\eta^{2\,n}\ll 1$ at very high $n$. Thus, the prospects for detecting the presence of the CNL is best for very low photon number (e.g. $n\in\{1,2\}$, and suggests the use of photon number resolving detectors, as suggested in \cite{eHOM_PRA:2022}, which can now detect and number-resolve up to 100 individual photons \cite{eaton_pfister_100_photons:2023}. 
 \section{Summary and Conclusion}\label{sec:Conclusion}
 The two photon HOM effect with input state $\ket{11}_{12}$ to a lossless 50:50 beam splitter with detection on the output coincidence state $\ket{11}_{12}$ reveals that the total quantum amplitude contains two terms, of equal magnitude but of opposite sign, hence summing to zero. By default of the choice of the input state, the two amplitude contribution diagrams in question are automatically ``mirror-images" of each other, namely  diagrams/amplitudes which swap the number of photon transmitted/reflected  with the number reflected/transmitted respectively, for both modes. 
 
 What the extended HOM effect unveils is that for higher order odd-odd dual Fock inputs 
 $\ket{n,m}_{12}$, the total amplitude for the output coincident state 
 $\ket{\tfrac{n+m}{2}, \tfrac{n+m}{2}}_{12}$ consists of a sum of pairs of multi-photon mirror-image component amplitude  diagrams  which have equal magnitude, but opposite sign. Thus, the total amplitude cancels pairwise on multiple  mirror-image amplitude contributions. In a sense, this is a series of  HOM-like effects on each of the $\thalf(n+1)$ mirror-image pair amplitude contributions.
 
 For even-even dual Fock inputs $\ket{n,m}_{12}$, the total amplitude for the output coincident state 
 contains an odd number of terms consisting of $\thalf\,n$ mirror-image pairs, and a lone un-paired ``middle" term. However, since this time $n$ is even, the mirror-image pair amplitude contributions are equal in magnitude with the \tit{same} sign, and so constructively (vs destructively) interfere. Regardless, the un-paired middle term is non-zero, so the net result is that for a 50:50 BS, there will always be a non-zero constructive interference, when $n$ is even.
 
 The implication of this result, is that for any odd-parity state (consisting of only odd number of photons) entering the mode-1 input port, then regardless of the state entering the mode-2 input port, be it pure or mixed, there will always be a central nodal line of zeros (destructive interference) along the the diagonal of the joint output probability distribution. This the statement of extended HOM  effect\cite{eHOM_PRA:2022}.
 
 In this work, we have first explored this mirror-image pair cancellation diagrammatically, before proceeding to a general analytic proof for a lossless 50:50 BS. We considered the prospects for realization of the extended HOM effect on the $\ket{1,1}_{12}$ output coincidence state for the experimentally accessible Fock state/coherent state  input state $\ket{1,\beta}_{12}$. 
 Finally, we  explored the modification of the extended HOM effect due to imperfect detection efficiency in anticipation of an experiment using a single photon/coherent state input.
 In a forthcoming paper \cite{eHOM:Is:Odd:Full:In:Prep:2024} we explore the modifications of the extended HOM effect due to the frequency difference of the two input states, and a variable time difference between the output photon detections. 
 
 
The lesson of the HOM and extended HOM effects highlights the fundamental importance of the discreetness of photonic quanta (which can be observed by photon counting) in quantum interference effects, 
 and the power of the non-classicality of quantum states, for even a single photon, to have a measurable effect on a macroscopic classical-like state (the coherent state ``laser" $\ket{\beta}$).
 
\begin{acknowledgments}
Any opinions, findings and conclusions or recommendations expressed in this material are those of the author(s) and do not necessarily reflect the views of their home institutions.
\end{acknowledgments}

\flushleft{\bf Conflict of Interest Statement:}  The authors have no conflicts to disclose.

\appendix
\section{The role of the beam splitter}\label{app:BS}
The role of a lossless beam splitter has been discussed widely by many authors over the years in many textbooks and articles, \cite{Loudon:2000, Agarwal:2013,Ou_Book:2017,Gerry_Knight:2023,Galvez_BS:2002} most notably for our purposes Skaar \tit{et al.}  \cite{Skaar:2004}.
A lossless beam splitter is defined by its representation of the unitary transformation $U$ of the input mode creation  operators $a^\dag_{1,in}, a^\dag_{2,in}$ to the output modes $a^\dag_{1,out}, a^\dag_{2,out}$ via
 $a^\dag_{i,out} = U\,a^\dag_{i,in}\,U^\dag\equiv \sum_j S_{ij}\,a^\dag_{j,in}$.
Here, the boson creation operators raise the number of photons in the Fock number state basis $\ket{n}$ by one according to the usual quantum mechanical harmonic oscillator rules (dropping subscripts) $a^\dag\ket{n} = \sqrt{n+1}\,\ket{n+1}$. The corresponding annihilation of photons (lowering the number by one) is given by the adjoint operation $a\ket{n} = \sqrt{n}\,\ket{n-1}$. 
For a two-port device, such as a beam splitter, we label both the input and output modes as mode-1 and mode-2. In general, the output creation operators can be expressed as a linear combination of the input creation operators, which we express as (gathering both modes in a two-component column vector)
\be{app:BS:S:out:in:terms:of:in}
\hspace{-0.5in}
\vec{a}^\dag_{out}=
\left[
\begin{array}{c}
a^\dag_{1,out} \\
a^\dag_{2,out} 
\end{array}
\right] = 
U
\left[
\begin{array}{c}
a^\dag_{1,in} \\
a^\dag_{2,in}
\end{array}
\right] 
U^\dag 
=
\left[
\begin{array}{cc}
S_{11} & S_{12}\\
S_{21} & S_{22}
\end{array}
\right]
\,
\left[
\begin{array}{c}
a^\dag_{1,in} \\
a^\dag_{1,in}
\end{array}
\right]. 
\ee
Writing out \Eq{app:BS:S:out:in:terms:of:in} explicitly as
\bsub
\bea{aout:intermsof:ain}
a^\dag_{1,out} &=& S_{11}\,a^\dag_{1,in} +S_{12}\,a^\dag_{2,in}, \label{app:aout:intermsof:ain:line1}\\
a^\dag_{2,out} &=& S_{21}\,a^\dag_{1,in} +S_{22}\,a^\dag_{2,in}, \label{app:aout:intermsof:ain:line2}
\eea
\esub
we see that $S_{ij}$ is interpreted as the scattering of an input photon from mode-$j$ (the second index)
into the output mode-$i$ (the first index). 
Unitarity of the $U$ requires the unitarity of the transfer matrix $S$ which can be expressed as the requirement of the orthonormality of the columns (and rows) of $S$, yielding \cite{Skaar:2004,Galvez_BS:2002}
\be{app:unitarity:of:S}
|S_{11}|^2 + |S_{12}|^2 =1,\quad 
|S_{21}|^2 + |S_{22}|^2 =1, \quad
S_{11}\,S^*_{12} + S_{21}\,S^*_{22}=0.
\ee
As stated in the main text, the unitarity of $S$ is simply a manifestation of the preservation of the commutators between the output modes, given the commutators of the input modes.

Taking the absolute value of the last equation in \Eq{app:unitarity:of:S} and using the first two equations yields 
$|S_{11}| =|S_{22}|$ and $|S_{12}| = |S_{21}| = \sqrt{1-|S_{11}|^2}$. Finally, writing each complex amplitude as 
$S_{ij} = |S_{ij}|\,e^{i\theta_{ij}}$, the last equation in \Eq{app:unitarity:of:S} 
yields the phase condition\cite{Galvez_BS:2002}
\be{app:phase:condition}
\theta_{11}+\theta_{22}=\theta_{12}+\theta_{21}+\pi.
\ee
In \Eq{app:S:various:forms} we show several forms of the beam splitter scattering matrix $S$ found in the literature.
\be{app:S:various:forms}
\left[
\begin{array}{cc}
S_{11} & S_{12}\\
S_{21} & S_{22}
\end{array}
\right] \rightarrow
%
%
\left[
\begin{array}{cc}
t & r\\
r & -t
\end{array}
\right], \quad 
\left[
\begin{array}{cc}
t & i r\\
i r & t
\end{array}
\right],\quad 
\left[
\begin{array}{cc}
t & -r\\
r & t
\end{array}
\right], \quad 
%
%
\begin{array}{c}
t= \cos(\theta/2)\\
 r= \sin(\theta/2)
\end{array}, \quad
0\le \theta\le \pi.
\ee
Here, $t, r\in \mathbb{R}$ are the transmission and reflection coefficients, respectively, such that the beam splitter transmissivity is given by $T=t^2$, and reflectivity by $R=r^2$ such that $T+R=1$. The beam splitter ``angle" $\theta$ is often conventionally chosen to lie in the range $\theta\in[0,\pi]$ such that a balanced, 50:50 beam splitter is given by $\theta= \pi/2$, where $t=r=\tfrac{1}{\sqrt{2}}$.

The  first  (asymmetric) matrix after the arrow in \Eq{app:S:various:forms} 
is applicable to a beam splitter in which mode-1 propagates in a high index of refraction material, while mode-2 propagates in a low index material \cite{Born_and_Wolf:1986, Hecht:2002, Skaar:2004}, while the 
second complex-symmetric matrix (employed in the main text) is applicable, for example, to  a pair of optical fibers brought close together in order to form a BS. The last representation of $S$ in in  \Eq{app:S:various:forms}  real-symmetric matrix (orthogonal) $2\times 2$ rotation matrix, whose advantage is that one only needs to tract the minus sign appearing only in $S_{12}$ for a photon reflecting from input mode-2 to output mode-1. 

The important point to note here is that none of the complete destructive interference results of the HOM and extended HOM effects depend on a particular choice of phase. The choice of phase does however represent different physical implementations of the beam splitter employed, even though the different matrices $S$ are all unitarily equivalent. The choice of phase also effects the relative phases of the remaining output state of the beam splitter, but again by a suitable unitary transformation, such states can be transformed into one other (i.e. the state are related by an appropriate unitary transformation, and so are also unitarily equivalent).

\providecommand{\noopsort}[1]{}\providecommand{\singleletter}[1]{#1}%



\begin{thebibliography}{10}

\bibitem{HOM:1987}
C.~K. Hong, Z.~Y. Ou, and L.~Mandel, ``Measurement of subpicosecond time
  intervals between two photons by interference,'' {\em Phys. Rev. Lett.},
  vol.~59, p.~2044, 1987.

\bibitem{eHOM_PRA:2022}
P.~M. Alsing, R.~J. Birrittella, C.~C. Gerry, J.~Mimih, and P.~Knight,
  ``Extending the {H}ong-{O}u-{M}andel {E}ffect: the power of
  nonclassiciality,'' {\em Phys. Rev. A}, vol.~105, p.~013712, 2022.

\bibitem{Kim:2002}
M.~Kim, W.~Son, V.~Buzek, and P.~Knight, ``Entanglement by a beam splitter:
  Nonclassicality as a prerequisite for entanglement,'' {\em Phys. Rev. A},
  vol.~65, p.~032323, 2002.

\bibitem{Ou:1996}
Z.~Ou, ``Quantum multiparticle interference due to a single photon,'' {\em
  Quant. and Semiclass Opt.}, vol.~8, p.~315, 1996.

\bibitem{Ou:2007}
Z.~Ou, {\em Multi-photon Quantum Interference}.
\newblock New York: Springer-Verlag US, 2007.

\bibitem{Ou_Book:2017}
Z.~Ou, {\em Quantum {O}ptics for {E}xperimentalists, (Chap. 6.2.2, p162 and
  8.3.2, p246)}.
\newblock Singapore: World Scientific, 2017.

\bibitem{Pan:2012}
J.~Pan, Z.~Chen, C.~Lu, and H.~Weinfurter, ``Multiphoton entanglement and
  interferometry,'' {\em Revs. Mod. Phys.}, vol.~84, p.~777, 2012.

\bibitem{Dakna:1997}
M.~Dakna, T.~Anhut, T.~Opatrny, L.~Kn{\"o}ll, and D.~Welsh, ``Generating
  schrodinger cat-like states by means of conditional measurements of a beam
  splitter,'' {\em Phys. Rev. A}, vol.~55, p.~3184, 1997.

\bibitem{Carranza:2012}
R.~Carranza and C.~C. Gerry, ``Photon-subtracted two-mode squeezed vacuum
  states and applications to quantum optical interferometry,'' {\em J. Opt.
  Soc. Am. B}, vol.~29, p.~2581, 2012.

\bibitem{Magana-Loaiza:2019}
O.~S. Maga{\~n}a-Loaiza, R.~Le{\'o}n-Montiel, A.~Perez-Leija, A.B., U’Ren,
  C.~You, K.~Busch, A.~Lita, S.~Nam, R.~Mirin, and T.~Gerrits, ``Multiphoton
  quantum-state engineering using conditional measurements,'' {\em NPJ Quant.
  Info.}, vol.~5, p.~80, 2019.

\bibitem{Dakna:1998}
M.~Dakna, L.~Kn{\"o}ll, and D.~Welsh, ``Photon-added state preparation via
  conditional measurement on a beam splitter,'' {\em Opt. Comm.}, vol.~145,
  p.~309, 1998.

\bibitem{Lvovsky:2002}
A.~Lvovsky and J.~Mlynek, ``Quantum-optical catalysis: Generating nonclassical
  states of light by means of linear optics,'' {\em Phys. Rev. Lett.}, vol.~88,
  p.~250401, 2002.

\bibitem{Bartley:2012}
T.~Bartley, G.~Donati, J.~Spring, X.~Min, M.~Barbieri, A.~Datta, B.~Smith, and
  I.~A. Walmsley, ``Multiphoton state engineering by heralded interference
  between single photons and coherent states,'' {\em Phys. Rev. A}, vol.~86,
  p.~043820, 2012.

\bibitem{Birrittella:2018}
R.~J. Birrittella, M.~El-Baz, and C.~C. Gerry, ``Photon catalysis and quantum
  state engineering,'' {\em JOSA B}, vol.~35, p.~1514, 2018.

\bibitem{Glauber:1995}
R.~Glauber, ``Letter to the editor,'' {\em Am. J. Phys.}, vol.~63, p.~12, 1995.

\bibitem{Bouchard:2021}
F.~Bouchard, A.~Sit, Y.~Zhang, R.~Fickler, F.~M. Miatto, Y.~Yao, F.~Sciarrino,
  and E.~Karimi, ``Two-photon interference: The {H}ong-{O}u-{M}andel effect,''
  {\em Rep. Prog. Phys.}, vol.~84, p.~012402, 2021.

\bibitem{eHOM:PNCs:note}
The non-diagonal curves that furcate all joint output probability distributions
  in \Fig{fig:eHOM:fig2:fig3:left:column:1photon:CS} were termed Pseudo Nodal
  Curves (PNC) and explored in detail by the authors of \cite{eHOM_PRA:2022}.
  The PNCs act as ``valleys" (local minima curves) where isolated zeros of
  $P(m_a, m_b|n)$ also sparsely occur. The more involved PNCs are not explored
  in this present work.

\bibitem{BMG:2012}
R.~J. Birrittella, J.~Mimih, and C.~C. Gerry, ``Multiphoton quantum
  interference at a beam splitter and the approach to {H}eisenberg-limited
  interferometry,'' {\em Phys. Rev. A}, vol.~86, p.~063828, 2012.

\bibitem{Lai:1991}
W.~Lai, V.~Bu{\u{z}}ek, and P.~Knight, ``Nonclassical fields in a linear
  directional coupler,'' {\em Phys. Rev. A}, vol.~43, p.~6323, 1991.

\bibitem{Campos:1989}
R.~Campos, B.~Saleh, and M.~Teich, ``Quantum-mechanical lossless beam spitter:
  {SU(2)} symmetry and photon statistics,'' {\em Phys. Rev. A}, vol.~40,
  p.~1371, 1989.

\bibitem{Loudon:2000}
R.~Loudon, {\em Quantum Theory of Light, 3rd ed.}
\newblock New York: Oxford University Press, 2000.

\bibitem{Galvez_BS:2002}
C.~H. Holbrow, E.~Galvez, and M.~E. Parks, ``Photon quantum mechanics and beam
  splitters,'' {\em Am. J. Phys.}, vol.~70, pp.~260--265, 2002.

\bibitem{Skaar:2004}
J.~Skaar, J.~Escartin, and H.~Landro, ``Quantum mechanical description of
  linear optics,'' {\em Am. J. Phys.}, vol.~72, p.~1385, 2004.

\bibitem{Agarwal:2013}
G.~S. Agarwal, {\em Quantum Optics}.
\newblock Cambridge: Cambridge University Press, 2013.

\bibitem{Gerry_Knight:2023}
C.~C. Gerry and P.~L. Knight, {\em Introductory Quantum Optics, 2nd Ed.}
\newblock Cambridge University Press,Cambridge, 2023.

\bibitem{Born_and_Wolf:1986}
M.~Born and E.~Wolf, {\em Principles of Optics, 6th Ed.}
\newblock New York: Pergamon Press, 1986.

\bibitem{Jackson:1999}
J.~D. Jackson, {\em Classical Electrodynamics, 3rd Ed.}
\newblock New York: John Wiley and Sons, Inc., 1999.

\bibitem{Hecht:2002}
E.~Hecht, {\em Optics, 4th Ed.}
\newblock San Francisco: Addison Wesley, 2002.

\bibitem{app:note:on:choice:of:S:in:HOM}
Note, that if we had used the leftmost (asymmetric) matrix representation of
  $S$ in \Eq{app:S:various:forms} we would have had the total HOM amplitude
  given by $A=(r)(r)+(-t)(t)\overset{t=r}{\to}0$. If we instead had used the
  last (real, annt-symmetric) representation of $S$, we would have obtained
  $A=(r)(-r)+(t)(t)\overset{t=r}{\to}0$. This demonstrates that the particular
  choice of the phases in the representation of $S$ does not effect the HOM
  complete destructive interference effect, as long as those phases satisfy the
  last (rightmost) equation in \Eq{app:unitarity:of:S}. The same holds true for
  the eHOM effect as well.

\bibitem{Legero_Rempe:2003}
T.~Legero, T.~Wilk, A.~Kuhn, and G.~Rempe, ``Time-resolved two-photon quantum
  interference,'' {\em Appl. Phys. B}, vol.~77, pp.~797--802, 2003.

\bibitem{Scully_Zubairy:1997}
M.~O. Scully and M.~S. Zubairy, {\em Quantum Optics, (Chap. 9)}.
\newblock Cambridge: Cambridge University Press, 1997.

\bibitem{eHOM:Is:Odd:Full:In:Prep:2024}
P.~M. Alsing, R.~J. Birrittella, C.~C. Gerry, J.~Mimih, and P.~Knight, ``The
  {E}xtended {HOM} effect for a balanced beam splitter,'' {\em (in
  preparation)}, 2024.

\bibitem{Boyd:1991}
R.~Boyd, {\em Nonlinear Optics}.
\newblock New York: Academic Press, 1991.

\bibitem{Ou:1996:Eq:4:note}
Ou \cite{Ou:1996} derived that $P(N_1,N_2) =
  \frac{N!}{N_1!\,N_2!\,2^{N+1}}\,(N_1-N_2)^2\, \delta_{N_1+N_2,N+1}$ for the
  input state $\ket{1,N}_{12}$. For a general mode-2 input state
  $\ket{1,\phi}_{12}$ with $\ket{\phi}_2 = \sum_N c_N \ket{N}_2$, one simply
  multiplies $P(N_1,N_2)$ by $|c_N|^2$. For a CS, $|c_N|^2 =
  e^{-|\beta|^2}\,\tfrac{|\beta|^2}{N!}$, which when multiplied by Ou's result
  yields the expression given in \Eq{P:N1:N2}.

\bibitem{Loudon:1983}
R.~Loudon, {\em Quantum Theory of Light, 2nd ed., (Sections: 6.6-6.8)}.
\newblock New York: Oxford University Press, 1983.

\bibitem{Knight:2002}
M.~S. Kim, W.~Son, V.~Bu$\check{z}$ek, and P.~L. Knight, ``Entanglement by a
  beam splitter: {N}onclassicality as a prerequisite for entanglement,'' {\em
  Phys. Rev. A}, vol.~65, p.~032323, 2002.

\bibitem{quantum:efficiency:note}
It is worth noting that the \textit{quantum efficiency} of the detector is
  given by (see Section 6.8 of Loudon \cite{Loudon:1983}) $\xi =
  \eta\,c\,\hbar\omega\,T/V = \eta\,\hbar\omega/\sigma$, where $\eta$ is the
  detector efficiency, $T$ is the integration time of the detector of photons
  of frequency $\omega$, and $V$ is a quantization volume. Here $\sigma =
  V/(T\,c)$ is the effective cross section for the detection process of the
  photons. The use of $\xi$ over $\eta$ allows one to incorporate the effect of
  very long integration times $T$. The net effect is that in the expression for
  $P_\eta(n)$ one replaces $\eta\to 1- e^{-\xi}$ so that $P_\xi(n; T) =
  \sum_{N=n}^\infty\binom{N}{n}\,(1- e^{-\xi})^n\,e^{-(N-n)\,\xi}\,P(N)$, which
  reduces to the expression given in the text when $\xi\ll 1$ and one can
  approximate $1- e^{-\xi}\approx \xi$ and $e^{-(N-n)\,\xi}\approx 1-(N-n)\,\xi
  \approx (1-\xi)^{N-n}$, with $\xi\to\eta$. In the limit of very large $\xi$
  one has $P_\xi(n; T)\overset{\xi\to\infty}{\longrightarrow}
  P(N)\,\delta_{N,n}$. For most detectors in current use with $\eta<1$, the
  expression in the text is adequate and appropriate.

\bibitem{eaton_pfister_100_photons:2023}
M.~Eaton, A.~Hossameldin, R.~J. Birrittella, P.~M. Alsing, C.~C. Gerry,
  H.~Dong, C.~Cuevas, and O.~Pfister, ``Resolution of 100 photons and quantum
  generation of unbiased random numbers,'' {\em Nature Photonics}, vol.~17,
  pp.~106--111, 2023.

\end{thebibliography}
\end{document}